\documentclass[aps, prc, tightenlines,superscriptaddress, letterpaper, amsmath, amssymb, preprintnumbers, floatfix, longbibliography, nofootinbib, hypertexnames=true]{revtex4-2}

\usepackage{graphicx, caption} 
\usepackage{dcolumn} 
\usepackage{bm} 
\usepackage{physics}
\usepackage[dvipsnames]{xcolor}
\usepackage{wrapfig}
\usepackage[font=small,labelfont=bf]{caption}
\usepackage{float}
\usepackage{tikz}
\usepackage{array}
\usepackage[normalem]{ulem}
\usepackage{tabularx}

\usepackage{orcidlink}

\usetikzlibrary{decorations.pathmorphing}
\usetikzlibrary {decorations.pathreplacing} 
\tikzset{snake it/.style={decorate, decoration=snake}}
\usetikzlibrary { datavisualization.formats.functions, datavisualization.polar }

\hypersetup{
    colorlinks=true,       
    linkcolor=blue,          
    citecolor=blue,        
    filecolor=blue,      
    urlcolor=blue           
}

\newcommand{\IQUS}{{InQubator for Quantum Simulation (IQuS), Department of Physics, University of Washington, Seattle, WA 98195, USA}}
\newcommand{\UNITN}{{Dipartimento di Fisica, University of Trento, via Sommarive 14, I–38123, Povo, Trento, Italy}}
\newcommand{\TIFPA}{INFN-TIFPA Trento Institute of Fundamental Physics and Applications,  Trento, Italy}

\begin{document}

\begin{figure}
\vskip -1.cm
\leftline{\includegraphics[width=0.15\textwidth]{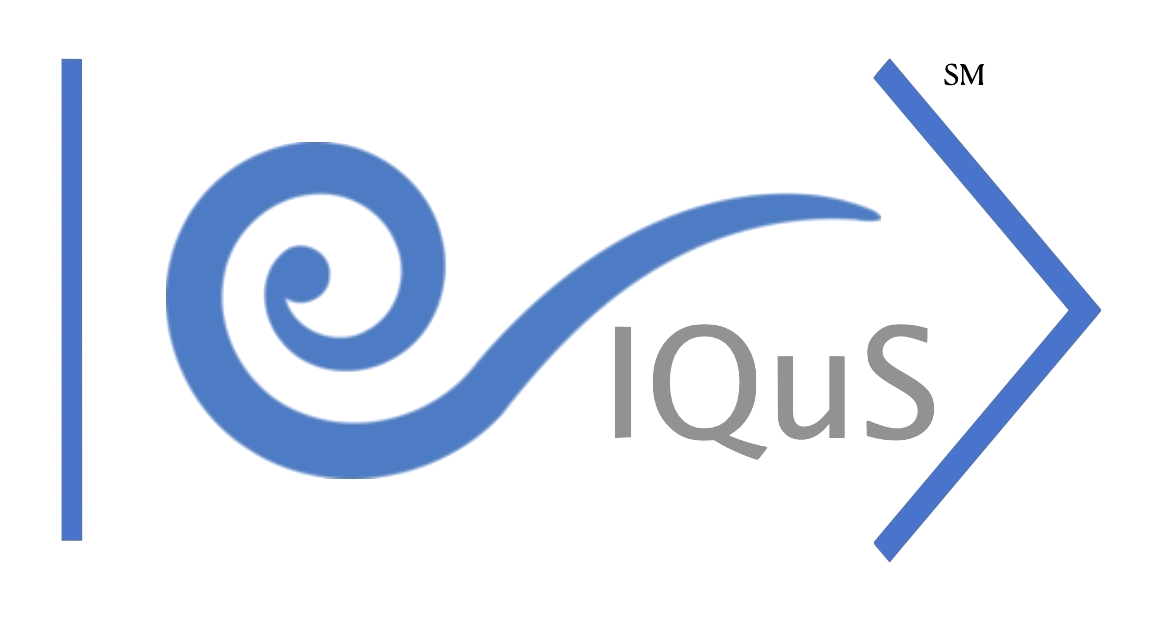}}
\vskip -0.5cm
\end{figure}

\title{Time Scales in Many-Body Fast  Neutrino Flavor Conversion}
\author{Ramya Bhaskar\,\orcidlink{0000-0003-2148-2590}}
 \email{rbhaskar@uw.edu} 
\affiliation{\IQUS}
\author{Alessandro Roggero\,\orcidlink{0000-0002-8334-1120}}
\email{a.roggero@unitn.it}
\affiliation{\IQUS}
\affiliation{\UNITN}
\affiliation{\TIFPA}
 \author{Martin J. Savage\,\orcidlink{0000-0001-6502-7106}}
 \email{mjs5@uw.edu} 
\thanks{On leave from the Institute for Nuclear Theory.}
\affiliation{\IQUS}

\preprint{IQuS@UW-21-069}
\date{\today}

\begin{abstract}
Time scales associated with many-body fast neutrino flavor conversions in core-collapse 
supernova are explored in the context of an effective two-flavor  model with axial symmetry.
We present a preliminary study of time scales obtained from a
linear stability analysis and from the distributions of Loschmidt echo crossing times 
(intimately connected to dynamical phase transitions in non-equilibrium systems)
determined by time evolution with the exact many-body Hamiltonian.
Starting from a tensor-product initial state describing systems of $N$ neutrinos, with $N/2$ electron-type and $N/2$ heavy-type, with uniform angular distributions, the Loschmidt echo crossing times, $t_{\mathcal{L}_{\times}}$, are found to exhibit two distinct time scales that are exponentially separated.  
The second peak structure at longer times, effectively absent for $N=4$, 
develops with increasing $N$. When re-scaled in terms of $\log t_{\mathcal{L}_{\times}}$,
the distributions are found to  become increasingly well described by the sum of two stable distributions. 
The distribution of Loschmidt echo crossing times differs somewhat from the results of the (numerical) linear stability analysis, which exhibits a peak at finite frequency and a second peak consistent with zero frequency.  The exact analysis suggests that the zero-frequency instability  manifests itself as a modest flavor-conversion time scale.
\begin{description}
\item[Keywords]
Neutrinos, Collective Flavor Conversions, Loschmidt Echos
\end{description}
\end{abstract}

\maketitle

\section{\label{sec:level1} Introduction}
\noindent
Neutrinos play a fundamental role in extreme astrophysical environments,
such as core collapse supernovae and neutron star binary mergers, 
where they, in part, 
determine the dynamical evolution of the explosion and set the ambient conditions for the ensuing nucleosynthesis, e.g., Refs.~\cite{Burrows_2021,mueller2019,wanajo2014production,hoffman1997nucleosynthesis,winteler2012magnetorotationally,Bruenn:2009ucj,Bruenn:2006oub}.
Important neutrino-matter processes active in these scenarios proceed through weak charged-current reactions, 
and can therefore be strongly affected by inhomogeneities in the flavor composition of the neutrino flux. 
It is well known that neutrinos can experience flavor oscillations in vacuum
(for a recent review, see Ref.~\cite{Workman:2022ynf}) but, for neutrinos with ${\rm MeV}$ energies, 
these occur over length scales that are typically large compared to the 
size of the supernova core.
In these  environments, 
flavor oscillations are modified by  charged-current scattering with charged leptons, leading to the MSW effect~\cite{mikheev1985resonance, wolfenstein1978neutrino}, 
and neutral-current neutrino-neutrino scattering which can lead to collective flavor oscillations~\cite{1987ApJ...322..795F,savage1991neutrino,pantaleone1992neutrino, pantaleone1992dirac,samuel1996bimodal,kostelecky1995,pastor2002flavor,duan2006collective,duan2006simulation} (see also Refs.~\cite{duan2010collective,CHAKRABORTY2016366,Tamborra2021} for reviews) which can dramatically reduce the oscillation length scales and thus possibly have important impact during the explosion~\cite{nagakura2021and,Nagakura2023}. The physics of collective flavor oscillations is typically discussed in terms of mean-field Boltzmann kinetic equations~\cite{sigl1993general,vlasenko2014,blaschke2016} which assume that neutrinos de-correlate between successive collisions among themselves. A better understanding of correlation effects beyond mean-field descriptions has been a topic of interest since the beginning of 
this research area~\cite{bell2003speed,friedland2003many,friedland2003neutrino,sawyer2004classical}, and has recently seen a resurgence of efforts~\cite{cervia2019,rrapaj2020exact,patwardhan2021spectral,roggero2021entanglement,roggero2021dynamical,xiong2022many,roggero2022entanglement,cervia2022,martin2022classical,lacroix2022,martin2023manybody,siwach233flav,martin2023eth} 
(see Ref.~\cite{patwardhan2023many} for a recent review, 
and also Refs.~\cite{PhysRevD.107.123004, johns2023neutrino} 
for recent discussions concerning
the many-body approaches)
aided in part by the use of 
available NISQ-era~\cite{Preskill_2018} quantum computers 
and by quantum information tools more generally~\cite{hall2021simulation,illa2022basic,amitrano23,illa2023multi,siwach23qc} 
(see also Refs.~\cite{Balantekin_2023,klco2022standard,bauer2023review} for recent reviews). 
A direct connection between the presence of unstable collective modes, 
whose amplitude grows exponentially at early times, 
with the presence of Dynamical Phase Transitions (DPT)~\cite{roggero2021dynamical,roggero2022entanglement}
has recently been shown.
This led to a proposal~\cite{roggero2022entanglement} of a 
unified description of unstable modes generated by asymmetries in the vacuum term, the so called "slow-modes" in the neutrino literature, 
with unstable modes generated by asymmetries in the flavor dependent angular distribution of neutrino velocities, the so called "fast-modes"
This identification was later shown to hold rigorously at the mean-field level~\cite{fiorillo2023slow}. In this work, we extend the previous analysis of Ref.~\cite{roggero2022entanglement}, that considered simple angular geometries, to the full multi-angle regime.

Neglecting the non-forward scattering processes between neutrinos and the external matter and neutrinos amongst each other, the Hamiltonian describing flavor evolution of neutrinos in dense environments contains three main contributions (see Ref.~\cite{duan2010collective} for a review): a one-body term responsible for neutrino flavor mixing in vacuum, a one-body term describing forward scattering between neutrinos and a static matter background (which gives rise to the MSW effect) and, finally, a two-body interaction term describing weak neutral-current forward scattering among neutrinos. In the two-flavor approximation, neutrinos can be described in terms of a $SU(2)$ flavor isospin, with the convention $\ket{\nu_e}=\ket{\uparrow}$ for the electron flavor and $\ket{\nu_x}=\ket{\downarrow}$ for the heavy flavor. In terms of flavor isospin, the Hamiltonian for $N$ neutrinos can then be written in the flavor basis as~\cite{balantekin2006neutrino, pehlivan2011invariants},
\begin{equation}
\label{eq:gen_ham}
\mathcal{H} =\sum\limits_{i = 1} ^{N}\frac{\Delta_m^2}{4E_i} \vec{B} \cdot \vec{\sigma_i} + \frac{\lambda}{2} \sum_{i=1}^N\sigma_i^z+ \frac{\mu}{2N} 
    		\sum\limits_{i<j} ^{N} (1-\cos(\theta_{ij}))\vec{\sigma_i}\cdot\vec{\sigma_j}
      \; ,
\end{equation}
where $\vec{\sigma_i} = (\sigma^x_i,  \sigma^y_i,  \sigma^z_i)$ is a vector of Pauli operators acting on the $i^{\rm th}$ neutrino and $E_i$ is the energy of the $i^{\rm th}$ neutrino. Neutrino masses are included by the squared-mass difference $\Delta_m^2 = m^2_2 - m^2_1$ and the mixing angle $\theta_{mix}$ in the normalized vector 
$\vec{B} = (\sin(2\theta_{mix}), 0, -\cos(2\theta_{mix}))$, reflecting the mass basis. Assuming a negligible density of heavy charged leptons, the coupling constant of the matter term is $\lambda=\sqrt{2}G_Fn_e$ with $G_F$ Fermi's constant and $n_e$ the electron-number density. The interaction term has a coupling constant $\mu=\sqrt{2}G_Fn_\nu$, with $n_\nu$ the total neutrino-number density, while $\theta_{ij}$ is the relative angle between the momenta of the $i^{\rm th}$ and $j^{\rm th}$ neutrinos, and encodes the geometry of the problem.

In this work, we are interested in the region close to the neutrino-sphere where the neutrino-neutrino coupling $\mu$ is much greater than the typical vacuum-oscillation frequency so that the first term in Eq.~\eqref{eq:gen_ham} can be neglected. In this regime, since the matter term commutes with the interaction term, for initial flavor states we can also neglect the presence of matter since it will not contribute to the ensuing flavor oscillation. Finally, we consider a geometry with axial symmetry with respect to some direction $\hat{z}$, which we take to be normal to a spherically symmetric neutrino-sphere, and average over the azimuthal angle $\phi$. The Hamiltonian then becomes,
\begin{equation}
\label{eq:axial_ham}
\mathcal{H}_{\nu\nu}=\frac{\mu}{2N} 
    		\sum\limits_{i<j} ^{N} (1-\cos(\theta_{i})\cos(\theta_{j}))\vec{\sigma_i}\cdot\vec{\sigma_j}
      \ =\ \frac{\mu}{2N} 
    		\sum\limits_{i<j} ^{N} \mathcal{J}_{ij}\vec{\sigma_i}\cdot\vec{\sigma_j}\;,
\end{equation}
with $\theta_i$ the polar angle of the $i^{\rm th}$ neutrino momentum, and where we introduce, for later convenience, the neutrino-neutrino coupling matrix $\mathcal{J}_{ij}$. In our numerical simulations, we use a forward-peaked uniform angular distribution with $\theta_i\in[-0.5,0.5]$.
We have verified that the conclusions reached with this choice 
of distribution remain unchanged
qualitatively when adopting a different angular range 
or other smooth non-uniform distributions (such as triangular or Gaussian).
As pointed out in previous works~\cite{martin2023manybody,martin2023eth}, the use of uniform grids of angles can lead to undesirable artificial degeneracies in the Hamiltonian's energy level. To prevent such (lattice) artifacts, the angles $\theta_i$ are randomly distributed uniformly on the interval $[0,0.5]$, 
and  ensemble calculations with different realizations are performed. As shown in Ref.~\cite{fiorillo2023slow}, the Hamiltonian in Eq.~\eqref{eq:axial_ham} is integrable, in the sense that it allows for an extensive number of conserved charges, and is therefore not able to thermalize the reduced one-body density matrices to a Boltzmann distribution~\cite{martin2023eth}. Nevertheless, it is a model commonly employed in studies of fast-flavor conversion~\cite{padilla2022neutrino1,fiorillo2023slow, PhysRevD.101.043009}, and provides a useful and interesting test case to analyze the time-scales associated to flavor oscillations in the many-body setting.

For the calculations described in this work,
an initial tensor-product
state of $N$ neutrinos given by
\begin{equation}
\label{eq:initial_state}
\ket{\Psi_0} = \otimes_{i=1}^{N/2}\ket{\nu_x}_i\otimes_{j=N/2+1}^{N}\ket{\nu_e}_j=\ket{\nu_x\dots\nu_x\nu_e\dots\nu_e}
\;,
\end{equation}
composed by $N/2$ electron neutrinos and $N/2$ heavy flavor neutrinos, is employed. This is a common initial state used in many previous works on many-body flavor evolution~\cite{roggero2021dynamical,roggero2021entanglement,lacroix2022,martin2022classical}, and has the advantage of enabling a focus on effects beyond mean-field as the state $\ket{\Psi_0}$ is invariant under mean-field evolution, and at the same time simplifies the analysis of DPT(s) in the model 
(see e.g., Ref.~\cite{roggero2021dynamical}). This is because a DPT is signalled by non-analiticities in the Loschmidt echo~\cite{heyl2013dynamical,heyl2018dynamical,vzunkovivc2018dynamical}, the probability of observing a time evolved state $\ket{\Psi(t)}=e^{-iHt}\ket{\Psi_0}$ in its initial condition $\ket{\Psi_0}$, presents itself as a crossing of two separate Loschmidt echoes associated with $\ket{\Psi_0}$ and its time-reversal counterpart~\cite{heyl2014dynamical}. This makes the numerical analysis easier to carry out than in the case where the Loscmidt echo shows a zero instead (as in Ref.~\cite{roggero2022entanglement}).

In this work,  Sec.~\ref{sec:lsa} presents a linear stability analysis of the collective modes supported by the axially symmetric Hamiltonian in Eq.~\eqref{eq:axial_ham} with the initial flavor state $\ket{\Psi_0}$ from Eq.~\eqref{eq:initial_state}. In Sec.~\ref{sec:losch}, a more detailed characterization of Dynamical Phase Transitions in terms of the Loschmidt echo is presented, specializing the discussion to the model considered in this work. 
In Sec.~\ref{sec:Analysis},  the results obtained for the Loschmidt echoes crossing time for an ensemble of neutrinos systems with sizes ranging from $N=4$ to $N=14$ are presented, along with a statistical analysis of the observed time scales. 
We summarize our findings  in Sec.~\ref{sec:summary},
and comment on potential directions of future research.

\section{Linear Stability Analysis}
\label{sec:lsa}
In order to shed  light on the expected collective modes present  in the system, we use the standard tool of linear stability analysis~\cite{banerjee2011,izaguirre2017fast,Airen_2018} (see also Ref.~\cite{roggero2022entanglement}). This starts with the evolution equations for the expectation values $\langle\vec{\sigma}_i\rangle$ within the mean-field approximation. For our model Hamiltonian Eq.~\eqref{eq:axial_ham}, these are
\begin{equation}
\frac{d}{dt}\langle\vec{\sigma}_i(t)\rangle=\frac{\mu}{N}\sum_{j\neq i}\left(1-\cos(\theta_i)\cos(\theta_j)\right)\langle\vec{\sigma}_j(t)\rangle\wedge\langle\vec{\sigma}_i(t)\rangle\;.
\end{equation}
For initial states aligned along the z direction and up to linear order in deviations from the initial condition, the third components $\langle\sigma^z_i\rangle$ remain static, 
while
the transverse components evolve via
\begin{equation}
\label{eq:lin_eom}
i\frac{d}{dt}\langle\sigma_i^\pm(t)\rangle=\frac{\mu}{N}\sum_{j\neq i}\left(1-\cos(\theta_i)\cos(\theta_j)\right)\left[\langle\sigma_j^z\rangle\langle\sigma_i^\pm(t)\rangle-\langle\sigma_i^z\rangle\langle\sigma_j^\pm(t)\rangle\right]=\sum_{j=1}^N\mathcal{M}_{ij}\langle\sigma_j^\pm(t)\rangle\;,
\end{equation}
where  $\sigma_i^\pm=(\sigma_i^x\pm i\sigma_i^y)/2$,  
decoupling the remaining two equations. 
The matrix $\mathcal{M}$ appearing in the right hand side is
\begin{equation}
\mathcal{M}_{ij} = \delta_{ij}\left[\frac{\mu}{N}\sum_{k\neq i}\left(1-\cos(\theta_i)\cos(\theta_k)\right)\langle\sigma_k^z\rangle\right]-(1-\delta_{ij})\frac{\mu}{N}\left(1-\cos(\theta_i)\cos(\theta_j)\right)\langle\sigma_i^z\rangle\;,
\end{equation}
which is manifestly not symmetric. Since the equations are linear, the solutions can be expressed as linear combinations of the $N$ fundamental solutions,
\begin{equation}
s^{(n)}_i(t)=\psi^{(n)}_i e^{-i\omega_n t}\;,
\end{equation}
where $\psi{(n)}_i$ is the component of the $n$-th eigenvector of $\mathcal{M}$ of the $i$-th neutrino and $\omega_n$ the corresponding eigenvalue. The eigenvalues can be either real or complex conjugate pairs, the latter case indicating the presence of unstable collective modes whose amplitude can grow exponentially in time. We will use the notation $\omega=\Omega+i\Gamma$, with real $\Omega$ and $\Gamma$, to denote the real and imaginary part of the eigenvalues. 

\begin{figure}[!tb]
		\includegraphics[width=\textwidth]{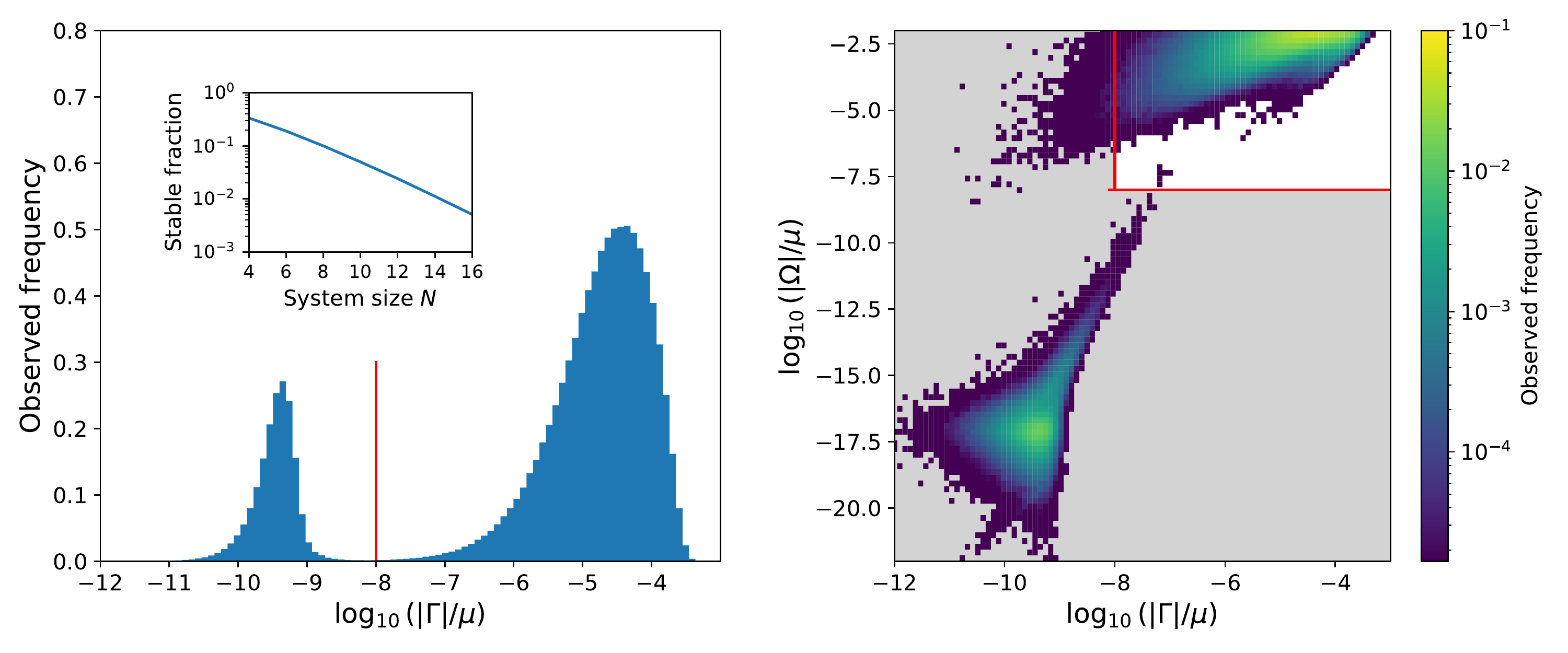}
\caption{ \raggedright The left panel shows a histogram of the observed frequency  of eigenvalues with a given imaginary component $\Gamma$ (in units of $\mu$) for a system with $N=14$, and using $10^6$ independent realizations of the angles $\theta_j$. The inset in the left panel shows the fraction of stable eigenvalues, with imaginary component $\Gamma=0$, as a function of system size. The right panel shows the observed frequency of unstable eigenvalues with a given magnitude of the real part $\Omega$ and imaginary part $\Gamma$ for the same system. In both panels the presence of modes with small values of $|\Gamma|$ and $|\Omega|$ is a numerical artefact, with the correct mode at zero frequency. We indicate the boundary of the region where numerical errors become important with red lines in both plots and on the right panel we show it also as a grey area.
   } 
   \label{fig:lsa_results}
	\end{figure}

In order to 
perform the linear stability analysis associated with the collective oscillation spectrum in this model,  
calculations of the eigenvalues for systems of even size from $N=4$ to $N=16$ were performed, using angles $\theta_i$ sampled uniformly in $[0,0.5]$. 
For each system size, 
we performed $10^6$ calculations corresponding to different random samples of the angles,  starting in the initial state from Eq.~\eqref{eq:initial_state}. 
The results of these calculations are displayed in Fig.~\ref{fig:lsa_results}. There are several points to note from these results. First, as shown in the inset on the left panel of Fig.~\ref{fig:lsa_results}, the fraction of stable eigenvalues (with imaginary component $\Gamma=0$) decreases exponential with the system size. This means that for large systems of neutrinos, the model described by Eq.~\eqref{eq:axial_ham} and with initial state $\ket{\Psi_0}$ from Eq.~\eqref{eq:initial_state} will contain unstable collective modes with high probability. This is in contrast to more geometrically symmetric systems such as the case of bipolar oscillations~\cite{roggero2021entanglement,martin2022classical,lacroix2022} or of fast oscillations in systems with three neutrino beams~\cite{roggero2022entanglement}. A somewhat unexpected result is the presence of two distinct types of unstable modes, as shown by the left main plot in the left panel of Fig.~\ref{fig:lsa_results}, 
which shows the observed frequency of occurrence (i.e., a histogram normalized so that the sum over the bin values times the bin width is unity) of unstable modes of different magnitudes of the imaginary part $\Gamma$ for $N=14$. The same behavior is observed for all values of $N$ examined. These two classes of modes are characterized by typical growth rates of either $\Gamma\approx 10^{-4}-10^{-5}\mu$ or $\Gamma\approx 10^{-9}-10^{-10}\mu$. In order to shed light on the structure of these two classes of modes, we present in the right panel of Fig.~\ref{fig:lsa_results} a heat map, for the same ensemble of systems used in the main plot of the left panel, showing the observed correlation between the magnitude of the real part $\Omega$ and imaginary part $\Gamma$ of the unstable modes. 
Unstable modes with large growth rates $\Gamma$ are associated with large real frequencies $\Omega$ and, conversely, modes with a small growth rate also have typically very small real frequencies. The observed values of the eigenvalues with small imaginary part are found to be strongly affected by floating-point errors, and should be regarded as eigenvalues with $\omega=0$.

The importance of considering both the real and imaginary part of unstable modes in the context of collective neutrino oscillations was recently pointed out in Ref.~\cite{padilla2022neutrino1}, 
where it was shown that, for a class of angular and flavor distributions, the oscillation depth quantifying the magnitude of flavor conversion was directly linked to the ratio $|\Gamma/\Omega|$: 
small values lead to a small oscillation depth,
while, conversely, large value  can lead to complete flavor inversion. 
In the present work, we use similar techniques (see also Ref.~\cite{fiorillo2023slow}) to provide a more concrete characterization of the observed two classes of modes in this model. We start by rewriting the linearized equations of motion Eq.~\eqref{eq:lin_eom} as,
\begin{equation}
\begin{split}
i\frac{d}{dt}\langle\sigma_i^\pm(t)\rangle&=\frac{\mu}{N}\sum_{j=1}^N\left(1-\cos(\theta_i)\cos(\theta_j)\right)\left[\langle\sigma_j^z\rangle\langle\sigma_i^\pm(t)\rangle-\langle\sigma_i^z\rangle\langle\sigma_j^\pm(t)\rangle\right]\\
&=\mu\left[S_0^z\langle\sigma_i^\pm(t)\rangle-\langle\sigma_i^z\rangle S^\pm_0(t)-\cos(\theta_i)S^z_1\langle\sigma_j^\pm(t)\rangle+\cos(\theta_i)\langle\sigma_i^z\rangle S_1^\pm(t)\right]
\ \ ,
\end{split}
\end{equation}
where the moments (equivalent to the vectors $\mathbf{D}_n$ in Refs.~\cite{padilla2022neutrino1,fiorillo2023slow})
\begin{equation}
\vec{S}_n(t)=\frac{1}{N}\sum_{j=1}^N \cos(\theta_j)^n \langle \vec{\sigma}_j(t)\rangle
\;,
\end{equation}
have been introduced.
We note that $\vec{S}_0$ is a conserved quantity owing to the global $SU(2)$ symmetry of the Hamiltonian $\mathcal{H}_{\nu\nu}$ in Eq.~\eqref{eq:axial_ham}
(a property maintained at the mean-field level),
and  $S^z_n$ are also conserved in the linear regime, as the third component of the flavor isospins is not dynamical. Furthermore, for our initial condition $\ket{\Psi_0}$ in Eq.~\eqref{eq:initial_state}, 
$S^\pm_0=0$ since the initial state has only components along $z$ and also $S_0^z=0$. 
Using these constraints, we finally have
\begin{equation}
\label{eq:eom}
\left[i\frac{d}{dt}+\mu\cos(\theta_i)S^z_1\right]\langle\sigma_i^\pm(t)\rangle=\mu\cos(\theta_i)\langle\sigma_i^z\rangle S_1^\pm(t)
\ \ .
\end{equation}
By noting that the right hand side is proportional to $\cos(\theta_i)\langle\sigma^z_i\rangle$, the solutions take the general form~\cite{padilla2022neutrino1,fiorillo2023slow},
\begin{equation}
\label{eq:pm_sol}
\langle\sigma_i^\pm(t)\rangle=\frac{A_\omega\cos(\theta_i)\langle\sigma^z_i\rangle}{\omega+\mu\cos(\theta_i)S^z_1}e^{-i\omega t}\;,
\end{equation}
with coefficient $A_\omega$ independent of $i$, 
and $\omega$ corresponds to the eigenvalues of the matrix $\mathcal{M}$ from Eq.~\eqref{eq:lin_eom}. 
Inserting these solution into Eq.~\eqref{eq:eom}, 
together with the condition $S^\pm_0=0$,
the eigenvalues satisfy,
\begin{equation}
1=\frac{\mu}{N}\sum_{j=1}^N\frac{\cos(\theta_j)^2\langle\sigma^z_j\rangle}{\omega+\mu\cos(\theta_j)S^z_1}\quad\quad\text{and}\quad\quad0=\frac{\mu}{N}\sum_{j=1}^N\frac{\cos(\theta_j)\langle\sigma^z_j\rangle}{\omega+\mu\cos(\theta_j)S^z_1}\;.
\end{equation}
Two types of solutions can be identified:
a uniform solution corresponding to $\omega=0$ with amplitudes essentially independent of the propagation direction $\theta_i$, 
and coherent solutions corresponding to higher frequencies and for which the eigenfunctions in Eq.~\eqref{eq:pm_sol} have amplitudes that are correlated with the propagation direction. 
The presence of a collective mode with $\omega=0$ has already been pointed out~\cite{fiorillo2023slow}, but it's structure was connected to the presence of amplitudes with $\cos(\theta_j)=0$ which we do not have. 
For our initial conditions however, this mode is still physical thanks to the fact that $S_0^z$ vanishes.
 
\section{Loschmidt Echoes}
\label{sec:losch}
\noindent
In this section, we provide a brief introduction to quantum quenches, the Loschmidt echo, and how Loschmidt echos can characterize DPTs via non-analyticities in Loschmidt rate functions. For a brief summary of the computational methods used for the simulations of  our neutrino system Eq.~\eqref{eq:axial_ham} and a generalization of Loschmidt echo-defined DPTs for our system, see section \ref{sec:Analysis}.

When studying real-time dynamics of an initial state $\ket{\Psi(t=0)}$
generated by a Hamiltonian $H$,
\begin{equation}
\ket{\Psi(t)} = e^{-itH}\ket{\Psi(0)},
\end{equation}
it is convenient to describe the evolution as a two step process: for time $t < 0$, 
the system is prepared in the ground-state of some initial Hamiltonian $H_\mathcal{I}$ and allowed to evolve trivially under the action of $e^{-itH_\mathcal{I}}$; 
at time $t=0$, the Hamiltonian 
is instantaneously changed to $H$, and 
the system is evolved forward in time.
This process of instantaneously changing the Hamiltonian $H_\mathcal{I} \rightarrow H$ at some $t=0$ is commonly called a ``quantum quench". ``Quenching" a system can include anything from suddenly turning on a $\vec{B}$ field, to altering some aspect of the system's geometry. For a review of quenches and recent theoretical advances, see Refs.~\cite{gogolin2016equilibration, mitra2018quantum}.
In this setting, a DPT is associated with the non-analytic behavior of a suitably defined out-of-equilibrium free energy. As initially proposed in Ref.~\cite{heyl2013dynamical}, the Loschmidt Echo $\mathcal{L}(t)$~\cite{gorin2006dynamics} can serve such a purpose. It is defined as
\begin{equation}
\label{eq:losch_echo}
\mathcal{L}(t)=\left|\langle \Psi(t)\vert\Psi(0)\rangle\right|^2=\left|\langle \Psi(0)\lvert e^{-itH}\rvert\Psi(0)\rangle\right|^2=e^{-N\lambda(t)}\;,
\end{equation}
and can be interpreted as probing the likelihood that the system returns to the initial state $\ket{\Psi(0)}$ after the quantum quench. The Loschmidt rate function $\lambda(t)$ serves the function of a intensive free-energy in this out-of-equilibrium process. In this setting, a DPT can be detected by the formation of a zero in $\mathcal{L}(t)$, which corresponds to a divergence in the rate $\lambda(t)$, as the system reaches the thermodynamic limit~\cite{heyl2013dynamical,heyl2018dynamical}.

\begin{figure}[t]
		\includegraphics[width=85mm,scale=0.7]{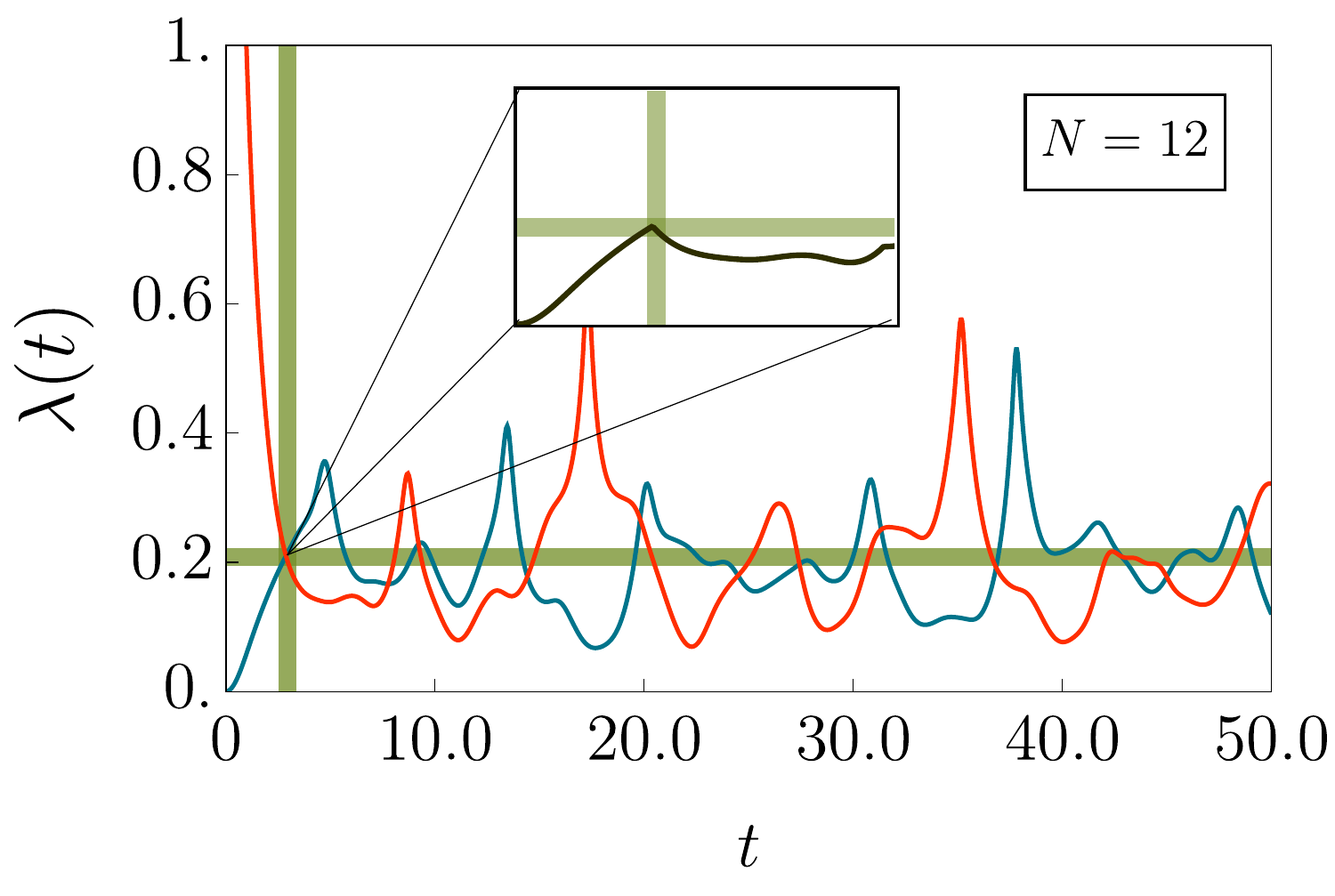}\qquad
		\includegraphics[width=85mm,scale=0.7]{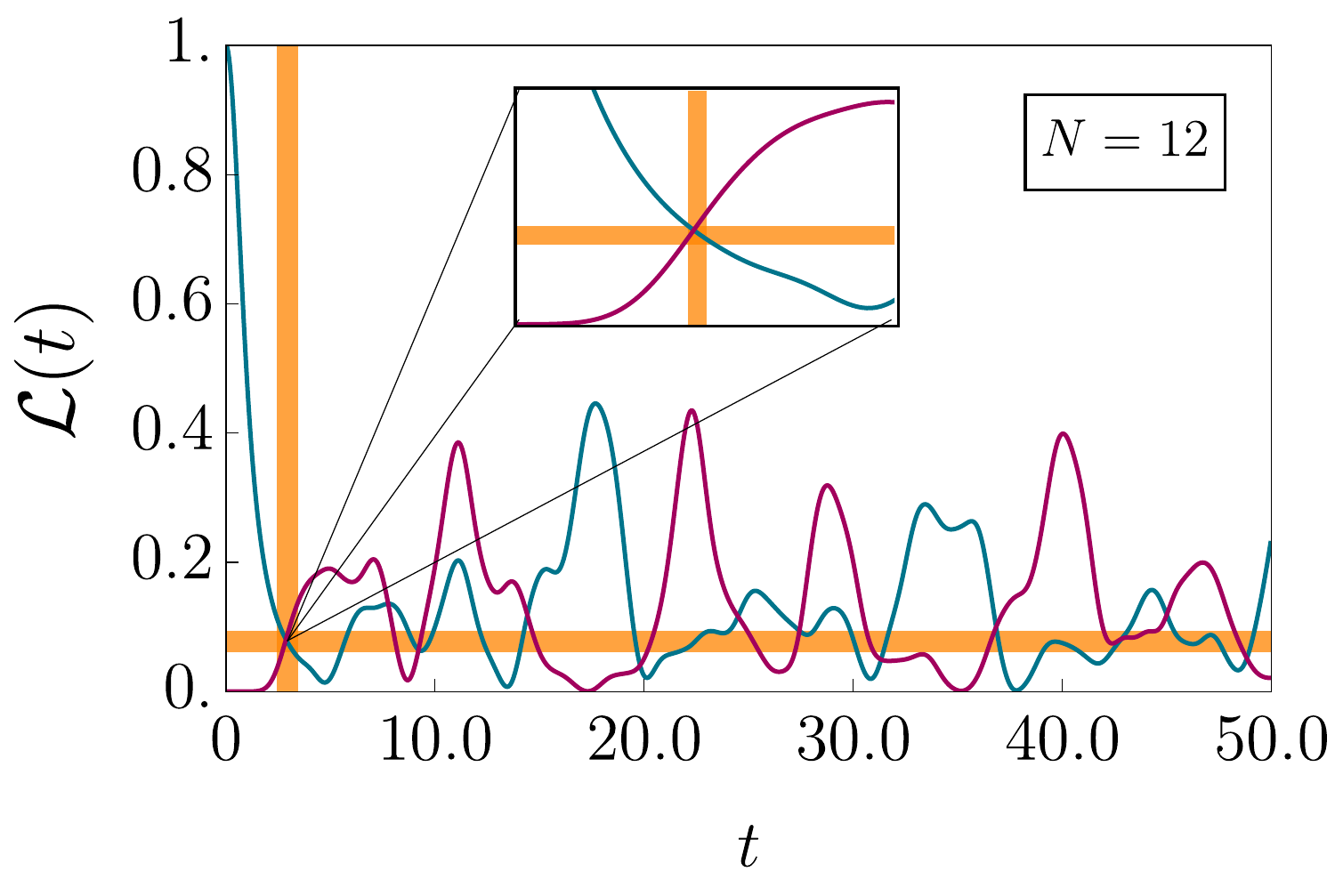}
\caption{ \raggedright 
The left panel shows $\lambda_{\beta,\beta}(t)$ (red curve) 
and  
$\lambda_{\alpha,\beta}(t)$ (blue curve)
as a function of time in units of $\mu^{-1}$. 
The thick green lines mark the  time when 
the rate functions first cross.
The inset zooms in on the lowest-$\lambda$'s first (non-analytic) cusp.
The right panel shows
$\mathcal{L}_{\beta,\beta}(t)$ (teal curve)  and  
$\mathcal{L}_{\alpha,\beta}(t)$ (purple curve). 
Orange lines mark the first crossing time, $t_{\mathcal{L}_{\times}}$,
in units of $\mu^{-1}$, with the inset focusing on the evolution around $t_{\mathcal{L}_{\times}}$.
   } 
   \label{fig:losch_echo}
	\end{figure}

The discussion of DPTs becomes more involved when the initial ground-state is degenerate, as in our model. Here we present the required generalization proposed in Ref.~\cite{heyl2014dynamical}, and used already in the collective neutrino oscillation setting in Ref.~\cite{roggero2021dynamical}.
Assume that $H_\mathcal{I}$ has two degenerate ground states: $\ket{\Psi_\alpha}$ and $\ket{\Psi_\beta}$, and we initialize our system $H_\mathcal{I}$ at $t\to-\infty$ to $\ket{\Psi_\beta}$. Then, at $t = 0$, we apply a quench to the system, where we instantaneously change $H_\mathcal{I} \rightarrow H$, and then time evolve $\ket{\Psi_\beta}$ under the quenched $H$. 
Two separate Loschmidt echoes can be defined,
\begin{equation}
\mathcal{L}_{\beta,\beta}(t) = \abs{\bra{\Psi_\beta}e^{-itH}\ket{\Psi_\beta}}^2\quad\quad\mathcal{L}_{\alpha,\beta}(t) = \abs{\bra{\Psi_\alpha}e^{-itH}\ket{\Psi_\beta}}^2
\ \ ,
		\label{eq:loschmidtprobabilitydef}	
\end{equation}
where the first, $\mathcal{L}_{\beta,\beta}(t)$, 
is analogous to the previous definition of the Loschmidt echo from Eq.~\eqref{eq:losch_echo}, 
while the second, $\mathcal{L}_{\alpha,\beta}(t)$, corresponds to the probability of transitioning to the orthogonal ground-state $\ket{\Psi_\alpha}$. 
The full out-of-equilibrium partition function is then defined as the sum of the two Loschmidt echoes~\cite{heyl2014dynamical}. Up to exponentially small corrections in the system size $N$, only one of the two rate functions will contribute to the free-energy,
and the overall rate can be defined as~\cite{heyl2014dynamical},
	\begin{equation}
		\label{eq:ratefunctionMinimumDominate}
		\lambda_{N}(t)=\min\left[\lambda_{\beta,\beta}(t), \lambda_{\alpha,\beta}(t)\right]\;.
	\end{equation}
With this definition, a non-analytic behavior, and thus a DPT, can present itself as a ``kink'' in the free energy when the dominant rate function switches between $\lambda_{\beta,\beta}(t)$ and $\lambda_{\alpha,\beta}(t)$ at some crossing time $t_{\mathcal{L}_{\times}}$. We display this behavior for a  system of neutrinos 
with $N=12$ neutrinos evolving under the Hamiltonian in Eq.~\eqref{eq:axial_ham} in Fig.~\ref{fig:losch_echo}. The left panel shows the two rate functions $\lambda_{\beta,\beta}(t)$ and $\lambda_{\alpha,\beta}(t)$ as a function of time,
with the inset focusing on the kink in $\lambda_N(t)$ formed by their crossing. The same behavior is of course paralleled by the crossing of the Loschmidt echoes, as shown in the right panel of Fig.~\ref{fig:losch_echo}.

Similar to the discussion of slow-modes in a neutrino system~\cite{roggero2021dynamical}, the Loschmidt echo crossing times that we compute are associated with the two ground states $\ket{\Psi_\alpha}$ and $\ket{\Psi_\beta}$,
\begin{eqnarray}
\ket{\Psi_\alpha }
& = & 
\bigg( \overset{N/2}{\underset{n=1}{\otimes}} \ket{\uparrow}_n  \bigg) \otimes \bigg( \overset{N}{\underset{m=N/2+1}{\otimes}} \ket{\downarrow}_m \bigg)
\ \ ,\ \ 
\ket{\Psi_\beta} \ = \  \bigg( \overset{N/2}{\underset{n=1}{\otimes}} \ket{\downarrow}_n \bigg) \otimes \bigg( \overset{N}{\underset{m=N/2+1}{\otimes}} \ket{\uparrow}_m \bigg)
\ \ ,
\label{eq:degenerate} 
\end{eqnarray}
where we took the state $\ket{\Psi_\beta}$ as the initial state for evolution at the quench, 
which is the same as the initial state $\ket{\Psi_0}$ from Eq.~\eqref{eq:initial_state} that is used in the linear stability analysis of Sec.~\ref{sec:lsa}.
For $N$ even, these are degenerate ground states of an initial 
Hamiltonian $H_\mathcal{I}$, 
\begin{equation}
	\label{eq:initialham} 
	H_{\mathcal{I}} =  \sum_{i=1}^{N/2}\sum_{j=N/2+1}^N  \sigma_i^z\sigma_j^z
 \ \ .
\end{equation}
At $t=0$, the system is quenched to evolve under the Hamiltonian $H_{\nu\nu}$ given
in Eq.~(\ref{eq:axial_ham}).
The initial state $\ket{\Psi_\beta}$ is evolved forward, with overlaps onto itself and onto
$\ket{\Psi_\alpha}$ computed, and  the first crossings of the Loschmidt echos, $\mathcal{L}_{\alpha,\beta}(t)$ and $\mathcal{L}_{\beta,\beta}(t)$,
are subsequently determined.

\section{Analysis of Loschmidt Echo Crossing Times}
\label{sec:Analysis}
\noindent
Ensembles of independent $\mathcal{J}_{ij}$ configurations 
with $N=4, 6, 8, 10, 12$ and $14$ neutrinos
were generated using Monte Carlo uniform sampling in $\theta_i$
for $0\le \theta_i \le 0.5$.
These provided ensembles of $H_{\nu\nu}$ defined in Eq.~(\ref{eq:axial_ham}).
The initial state in Eq.~(\ref{eq:degenerate})
was evolved forward in time for each member in an ensemble, and the 
times of the first crossing of the 
Loschmidt echos $\mathcal{L}_{\alpha,\beta}(t)$ and  $\mathcal{L}_{\beta,\beta}(t)$,
given in Eq.~(\ref{eq:loschmidtprobabilitydef}),
were determined.
For the systems with $N\le 10$, 
the systems could be evolved in time 
and the Loschmidt echos determined
exactly using {\tt Mathematica}, 
without the need for time decimation.
For larger systems (and the smaller systems for verification purposes),
matrix product states (MPS) were used to simulate dynamics 
via {\tt iTensor}’s {\tt Julia} package \cite{fishman2022itensor}
using Trotterization of the evolution operator,  
implemented via the Time Evolving Block Decimation (TEBD) 
method~\cite{paeckel2019time, vidal2004efficient, suzuki1976generalized},
with a tolerance of $10^{-9}$.
A maximum evolution time interval of $50 \left(2 N/\mu\right)$ was selected, 
with a Trotter step size of 
$\Delta t=0.05 \left(2 N/\mu\right)$~\footnote{
Step sizes of  $\Delta t  = [0.02, 0.07, 0.1  ] \left(2 N/\mu\right)$ were also profiled,
and furnished consistent results. 
}.
This was 
guided by previous numerical results~\cite{friedland2003many,friedland2003neutrino} 
which demonstrated collective flavor oscillations 
taking place on time scales typically ranging from 
$t_{slow} = \mu^{-1}\sqrt{N}$ to 
$t_{fast} = \mu^{-1}$log$(N)$.
To account for the increased complexity in the system introduced by 
our random selection of neutrino momenta, 
the maximum simulation time was increased  by $10\times$,
as similarly estimated in Ref.~\cite{roggero2022entanglement}. 

\begin{figure*}[t]
		\includegraphics[width=180mm,scale=1.0]{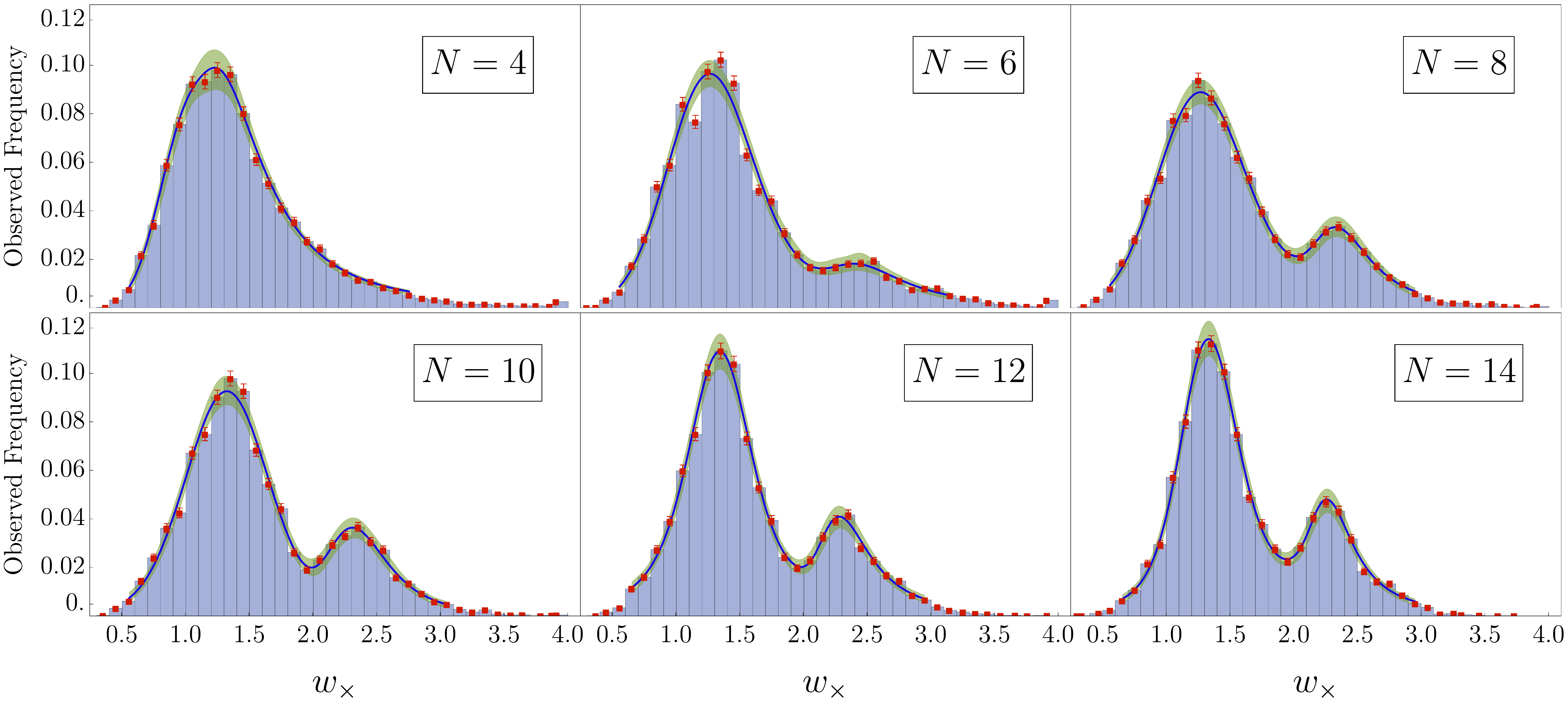}
		\caption{
  \raggedright
Histograms of the Loschmidt-echo crossing times for system sizes of 
$N = 4, 6, 8, 10, 12$ and $14$ neutrinos 
evolving under the dynamics described in the text.
They show the number of events in a given bin normalized to the total number of events, 
as a function of $w_{\times}$, defined in Eq.~(\ref{eq:wdef}). 
A bin width of $\delta w_{\times}=0.100$ is employed in each histogram.
All distributions were formed from $10.5$k independent samples of $\mathcal{J}_{ij}.$
The red points with error bars correspond to the height of each bin 
and uncertainty estimated by bootstrap re-sampling over the $10.5$k independent samples.
The dark-blue curves show the best fit  
of the sum of two stable distributions
(given in Eq.~(\ref{eq:fitforms}))
to the histograms,
with fit parameters and uncertainties given in 
Table~\ref{tab:BestFitParamTable}.
The $68\%$ confidence intervals associated with the fits are shown by the 
green-shaded regions.
}
  \label{fig:bimodalhistogramwide} 
\end{figure*}

The observed heavy tails of the obtained distributions, 
reminiscent of those found in lattice QCD calculations of baryon correlation functions
(see, e.g., Ref.~\cite{Wagman:2016bam}),
prompts us to change variables from 
$t_{\mathcal{L}_{\times}}$ to $w_{\times}$, where  
\begin{eqnarray}
   w_{\times} & = &  
 \log  t_{\mathcal{L}_{\times}}
   \ \ . 
   \label{eq:wdef}
\end{eqnarray}
Figure~\ref{fig:bimodalhistogramwide} displays the associated
histograms  as functions of 
$w_{\times}$ (with uniform bins width) for each $N$. 
As the systems are evolved forward in time up 
to a  maximum time, $T_{\rm max}$,
there are a small number of elements in the ensembles 
that do not experience a Loschmidt Echo crossing (during that time interval).
Qualitatively, 
for the sizes of ensembles we have generated,
the $N = 4$ system exhibits a single skewed distribution, 
with a heavy tail for increasing $w_{\times}$.
The $N>4$ systems show the emergence of a second peak on the shoulder of the first.

For small system sizes, the discreteness of the systems furnish ``sectors'' of Loschmidt echo crossing times, as suggested in Fig.~\ref{fig:medianJijwide}.  As a result, the distributions are not expected to be smooth functions for small $N$. 
However, 
they are expected to become smooth with increasing $N$.  
Therefore, fitting to smooth functions should describe  
results for large $N$, but become increasingly poor descriptions at small $N$.
In order to quantify these distributions,
the sum of two stable distributions are fit to each histogram,
in particular, type-1 stable distributions,
\begin{eqnarray}
    h(w) & = & 
    a_1 \mathcal{S}_1(\alpha_1, \beta_1, \mu_1, \sigma_1, w)
    \ +\ 
    a_2 \mathcal{S}_1(\alpha_2, \beta_2, \mu_2, \sigma_2, w)
    \ ,
    \label{eq:fitforms}
\end{eqnarray}
where
$a_i$ denote their amplitudes and are real numbers.
As defined in {\tt Mathematica},
the type-1 stable distribution is, 
\begin{eqnarray}
\mathcal{S}_1(\alpha, \beta, \mu, \sigma, x)
 & = & 
 e^{i\mu t 
 - |t\sigma|^\alpha \left(
 1-i \beta \tan \left({\pi\alpha\over 2}\right) \ {\rm sgn(t)} 
 \right)
 }
  \ : \ \alpha\ne 1
  \nonumber\\
   & = & 
 e^{i\mu t 
 - t\sigma \left(
 1+2 i {\beta\over \pi} \log |t|\  {\rm sgn(t)} 
 \right)
 }
 \ : \ \alpha = 1
\ \ ,
\end{eqnarray}
where $\mbox{sgn}(x)$ is the sign of $x$.
$\alpha $ is the stability parameter, 
$\beta $ is the skewness parameter, 
$\mu $ is the mean of the distribution, 
and $\sigma$ is twice the variance.
The results of fitting $h(w)$ in Eq.~(\ref{eq:fitforms}),
with a total of 10 fit parameters,
are shown in
Table~\ref{tab:BestFitParamTable},
including the uncorrelated weighted $\chi^2/{\rm dof}$ to indicate the goodness of fit.
\begin{table}[t]
\renewcommand{\arraystretch}{1}
\begin{tabularx}{0.95\textwidth}{c|cccccccccc|c}
 \hline
 $N$ 
 & $a_1$ 
 & $\alpha_1$ 
 & $\beta_1$ 
 & $\mu_1$ 
 & $\sigma_1$
 & $a_2$ 
 & $\alpha_2$ 
 & $\beta_2$ 
 & $\mu_2$ 
 & $\sigma_2$
 & $\chi^2/{\rm dof}$
  \\
 \hline\hline
4  & 0.0849(54) & 1.34(10) & 0.976(76) & 1.64(16) & 0.275(23) & 0.0231(29) & 0.892(43) & 0.621(76) & 0.35(31) & 0.313(34) & 2.03 \\
6  & 0.0840(35) & 1.842(86) & 0.914(61) & 1.333(35) & 0.2474(93) & 0.0184(11) & 1.049(20) & 0.389(28) & 4.32(66) & 0.358(22) & 6.54 \\
8  & 0.0793(29) & 1.866(81) & 0.918(41) & 1.326(37) & 0.2531(96) & 0.0231(11) & 1.175(62) & 0.468(22) & 2.76(16) & 0.226(11) & 1.69 \\
10 & 0.0785(27) & 1.854(86) & 0.0785(41) & 1.330(23) & 0.2397(91) & 0.0230(10) & 1.366(78) & 0.898(59) & 2.630(90) & 0.1893(88) & 3.56\\
12 & 0.0843(36) & 1.349(83) & 0.00277(15) & 1.346(21) & 0.2268(9) & 0.0237(12) & 1.121(57) & 0.613(35) & 2.97(27) & 0.1952(99) & 2.01\\
14 & 0.0771(31) & 1.447(72) & 0.213(12) & 1.389(25) & 0.1966(76) & 0.0283(15) & 1.079(48) & 0.313(16) & 2.80(27) & 0.1987(83) & 2.31\\
 \hline
\end{tabularx}
	\caption{
   \raggedright
   Best-fit parameters obtained from fitting two stable distributions,
 given in Eq.~(\ref{eq:fitforms}),  to the histogram distributions of 
 Loschmidt echo crossing times for systems with $N = 4, 6, 8, 10, 12, 14$ neutrinos, 
 as displayed in Fig.~\ref{fig:bimodalhistogramwide}.
 The histogram bin size is $\delta w_{\times}  = 0.100$. 
 All distributions were formed from $10.5$k independent samples of $\mathcal{J}_{ij}$.
 }
\label{tab:BestFitParamTable} 
\end{table}
Monte Carlo sampling was used to perform the fitting and in determining the $68\%$ CIs.
For each histogram,
the global minimum of the uncorrelated weighted $\chi^2/{\rm dof}$~\footnote{
The weights corresponded to the inverse variance of each histogram bin, determined by bootstrap re-sampling.} 
was found by a 
search in the 10-dimensional parameter space, first using a coarse lattice of points then
followed by a stochastic gradient descent.
The $68\%$ CI was  determined by sampling parameter space to identify the surface for which $\chi^2_{\rm min} \rightarrow \chi^2_{\rm min} + 11.55$, 
appropriate for a 10 parameter fit.
In some instances, a second minimum lay within this upper value of $\chi^2$,
giving outliers in the sampling.
Their impact was mitigated by using the median average deviation (MAD) 
of the parameter sample, 
with an appropriate re-scaling factor (computed to be 4.855(9)).
While the presented results use bin widths of $\delta w_{\times}  = 0.100$, 
a range of bin widths were examined $\delta w_{\times}  = 0.060, 0.065,...0.140$,
over which the resulting fit parameters were found to be consistent.

There are some interesting points to note from our results.
First is that the locations of the two peaks, $\mu_{1,2}$, become approximately 
independent of the number of neutrinos, and the variances also appear to be approximately stable.   
Further, 
the relative strength of the two distributions appears somewhat stable to increasing neutrino number.
The goodness-of-fit to two stable distributions depends on the number of neutrinos, as can be seen in the last column of  Table~\ref{tab:BestFitParamTable}.
For the systems with $N\ge 8$, there are two peaks in the histograms 
that are reasonably well reproduced by the sum of two stable distributions. 
However, as the $\chi^2/{\rm dof}> 1$ for each, 
this does not constitute a perfect description of the systems, as expected.
It is found to be a superior hypothesis to that of a single stable distribution for 
$N\ge 8$, which have $\chi^2/{\rm dof}> 10$.

It is enlightening to compare the results for the Loschmidt echo crossings with the 
results of the stability analysis that we presented above,
in particular the distributions of imaginary parts of the frequencies.
In both instances, there are two peaks in the distributions associated with the sampled ${\cal J}_{ij}$ matrices.  
The slower and faster modes correspond to the long-time scale and short-time scale Loschmidt echo crossings, respectively.
Interestingly, while the distributions associated with 
faster-modes and the shorter Loschmidt echo crossing times 
occur at the same $w_{\times}$, 
those associated with the slower modes 
(consistent with zero frequency)
are at a desparately larger value of 
$w_{\times}$ than the corresponding longer Loschmidt echo crossing times.
The well-separated slower- and faster-mode distributions 
differ from the 
overlapping distributions in the Loschmidt echo crossing times.
So while the linear instability analysis describes well the fast modes, 
it is only qualitatively reflective of the 
longer Loschmidt echo crossing times.

\section{Summary}
\label{sec:summary}
\noindent
Neutrinos play a key role in core-collapse supernova.
After creation in the core, neutrino flavor evolves in complex ways 
through interactions with matter, with other neutrinos, 
and because the mass eigenstates are combinations of the weak eigenstates.
Our work builds upon current research into the quantum evolution of model neutrino systems in the full many-body treatment. 
Starting with a tensor-product state of equal numbers of electron-type and heavy-type neutrinos, small- and modest-sized systems are evolved forward by the Hamiltonian describing neutrino-neutrino 
forward scattering, neglecting vacuum mixing. In this setting, the forward-scattering interactions with matter only contribute a global phase factor and are therefore neglected. 
Attention is paid to the time-scales for flavor conversion under this evolution, both 
through a linear stability analysis and also using the crossing of Loschmidt echos.
We consider Loschmidt echo crossings because of their relation to dynamical phase transitions in non-equlibrium systems.

Through Monte Carlo sampling over a uniform distribution in neutrino angles,
we generate ensembles of neutrino coupling matrices in the context of 
an axially-symmetric light-bulb model.
In our linear stability analysis, both zero-frequency and non-zero frequencies instabilities are found,
while two modes are observed in the crossings of the Loschmidt echoes, 
both with finite time-scales.
The locations of the peaks in the Loschmidt echo crossings appear independent of the number of neutrinos for $N\le 14$.
While the distributions for the $N=4,6$ systems reflect the underlying discreteness of the systems, we found that they become increasingly well described by smooth distributions,
in particular the sum of two stable distributions.
As a mean-field analysis identifies only one fast mode of flavor conversion for such systems, our fully quantum mechanical analysis has revealed the presence of a second mode.
Further work is required to better understand the impact of this second mode, and its connection to the  linear stability analysis. In future work, it would be interesting to understand the role played by the integrability of the axially-symmetric Hamiltonian, and whether a rich multi-peaked spectrum could be expected also in the chaotic multi-angle regime studied in Ref.~\cite{martin2023eth}.
In this context, the zero mode may potentially show it's presence in thermalizing systems as a quantum many-body scar, creating unusually long thermalization times.

\section{Acknowledgements}
\noindent
We thank the following colleagues for their valuable discussions, feedback, and suggestions throughout this work:  
Joe Carlson, 
Anthony Ciavarella, Vicenzo Cirigliano, Roland Farrell, Marc Illa, 
Luke Johns, 
Saurabh Kadam, David Kaplan, Natalie Klco,  Joshua D. Martin, Niklas Mueller,  Hersh Singh, Jesse Stryker, Francesco Turro, Xiaojun Yao, 
and the InQubator for Quantum Simulation (IQuS) for providing an intellectually stimulating environment. 
We would like to thank the QC4HEP Working Group at CERN and the Thrust-2 team in the Quantum Science Center (QSC)
(\url{https://qscience.org}), a National Quantum Information Science Research Center of the U.S.\ Department of Energy, 
that are focused on quantum information and simulation aspects of neutrino physics.
This work was supported in part by U.S. Department of Energy, Office of Science, Office of Nuclear Physics, InQubator for Quantum Simulation (IQuS)\footnote{\url{https://iqus.uw.edu/}}
under Award Number DOE (NP) Award DE-SC0020970 
via the program on Quantum Horizons: QIS Research and Innovation for Nuclear Science\footnote{\url{https://science.osti.gov/np/Research/Quantum-Information-Science}}
This work is also supported, in part, through the Department of Physics\footnote{\url{https://phys.washington.edu}} 
and the College of Arts and Sciences\footnote{\url{https://www.artsci.washington.edu}}
at the University of Washington.
We have made extensive use of Wolfram {\tt Mathematica}~\cite{Mathematica},
{\tt python}~\cite{python3,Hunter:2007}, {\tt julia}~\cite{Julia-2017},
{\tt jupyter} notebooks~\cite{PER-GRA:2007} 
in the {\tt Conda} environment~\cite{anaconda}.
This work was enabled, in part, by the use of advanced computational, storage and networking infrastructure provided by the Hyak supercomputer system at the University of Washington.\footnote{\url{https://itconnect.uw.edu/research/hpc}}

\appendix
\section{Statistics of the $\mathcal{J}_{ij}$ Neutrino Coupling Matrices}
\begin{figure}[b]
		\includegraphics[width=0.85\columnwidth]{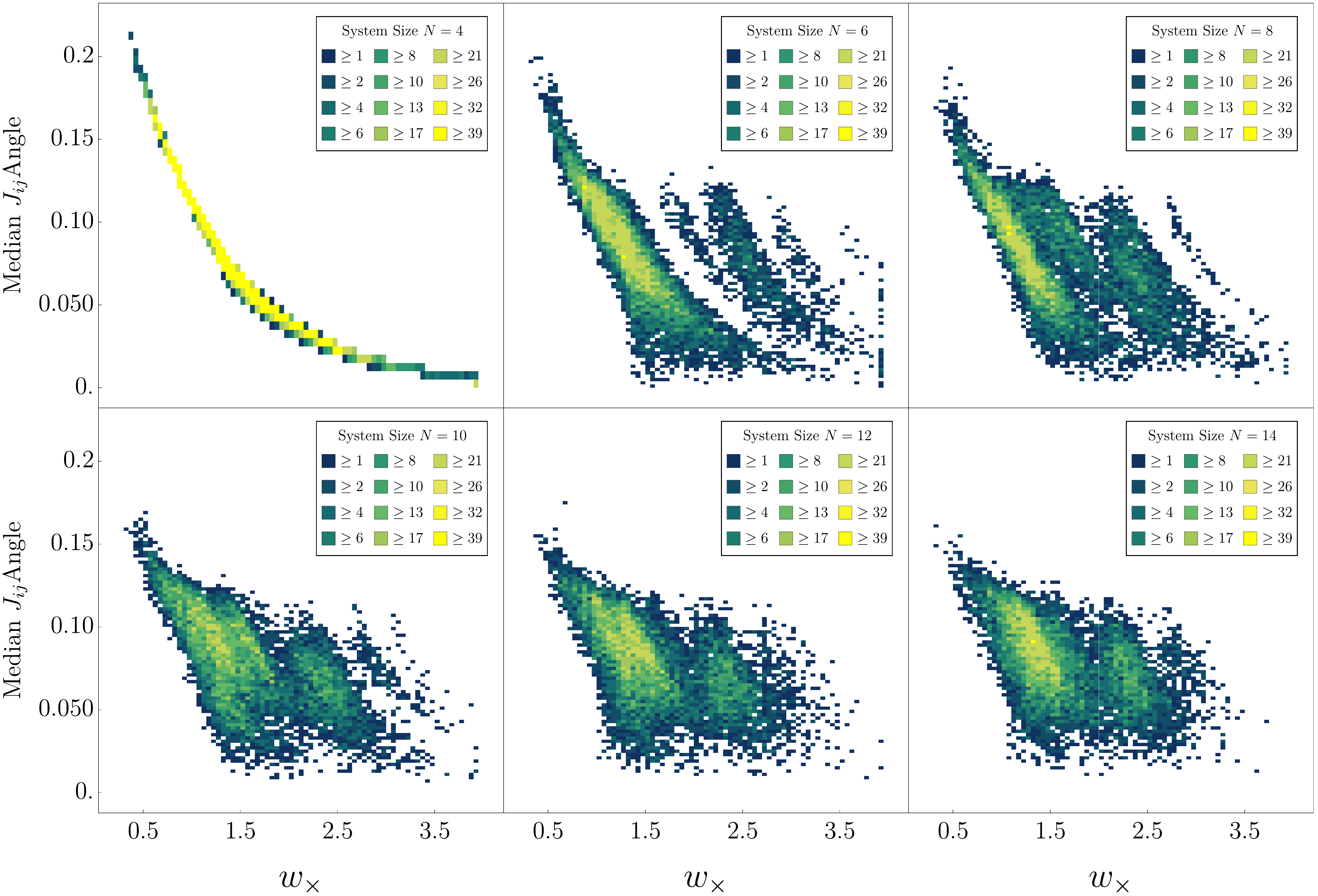}
		\caption{\label{fig:medianJijwide} 
			\raggedright  
   Scatter plots of the median $\mathcal{J}_{ij}$ for each element in the ensembles
   versus $w_\times$, defined in Eq.~(\ref{eq:wdef}),
   for the systems corresponding to the histograms shown in Fig.~\ref{fig:bimodalhistogramwide} and described in the text.
      }
\end{figure}

\noindent
To gain further insights into the presence of two peaks in the 
Loschmidt echo crossing times,
the medians of the $\mathcal{J}_{ij}$ coupling terms are examined. 
Figure \ref{fig:medianJijwide} shows the distributions of the median $\mathcal{J}_{ij}$ 
for each ensemble as a function of $w_\times$.
For $N=4$ there is essentially a 1-to-1 correspondence between the median $\mathcal{J}_{ij}$ and $t_{\mathcal{L}_{\times}}$, 
and a single peak in the histogram of crossing times follows naturally.
As the number of neutrinos increases, 
the distribution forms from an increasing number of bands, 
that merge into regions.
While two distributions are seen to emerge, 
it remains the case that the shortest times correspond to the largest median values.
While we have not explored this in detail, and leave it to future work, 
we anticipate that the band structure is related to the integrability of these systems,
and the overlap of the initial state onto different sectors.

\section{Further Results from the  Linear Stability Analysis}
\begin{figure}[b]
		\includegraphics[width=\textwidth]{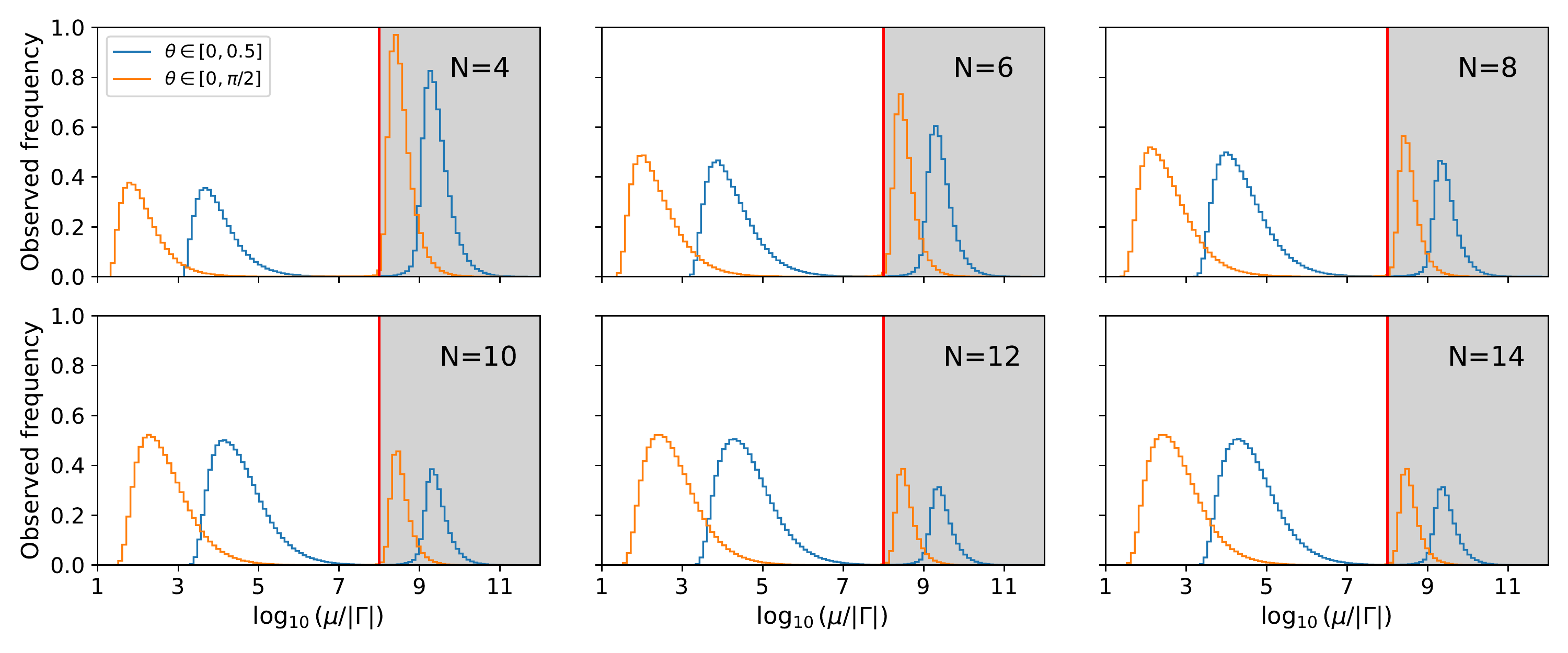}
\caption{ \raggedright 
Observed distribution of unstable time scales $\mu/|\Gamma|$ obtained through linear stability analysis of systems with size ranging from $N=4$ (top left) to $N=14$ (bottom right). Each panel shows two curves: the blue line corresponds to angles $\theta_j$ sampled uniformly in $[0,0.5]$ while the orange line corresponds to angles sampled on a wider interval $[0,\pi/2]$. Results showing time scales longer than $10^8 \mu^{-1}$ are strongly affected by numerical errors and correspond to zero-frequency modes. In order to highlight the region affected by numerical precision, we show it with a grey background and indicate its boundary as a vertical red line.
   } 
   \label{fig:lsa_add_results}
	\end{figure}

\noindent
In this Appendix, we show further results produced using the linear stability analysis described in Sec.~\ref{sec:lsa}. As noted there, the results are qualitatively independent of the size of the system and the choice of angular distribution: we always observe unstable modes with relatively large imaginary part $|\Gamma|$ and a second set of modes with much lower values of $|\Gamma|$ which, within numerical precision, correspond to the zero mode. The typical values of $|\Gamma|$, or equivalently the instability time scale $1/|\Gamma|$, are seen to be independent of the system size, while they are shifted by changing the angular range and shape. 
This is shown in Fig.~\ref{fig:lsa_add_results} for a range of system sizes, 
ranging from $N=4$ to $N=14$ (the same considered in the full simulation presented in Sec.~\ref{sec:Analysis}), 
and for two angular distributions where $\theta_j$ is distributed uniformly in $[0,0.5]$, shown as blue lines and corresponding to the results shown in Sec.~\ref{sec:Analysis}, and a wider distribution with $\theta_j$ uniform in the range $[0,\pi/2]$, shown as orange lines. 
As expected from the definition of the coupling matrix $\mathcal{J}_{ij}$ in Eq.~\eqref{eq:axial_ham}, the second set of results corresponding to a wider angular distribution are found to produce faster modes (on average). 
In order to more clearly highlight the region of time-scales affected by numerical errors,
it is shown by a grey background with a boundary indicated by  
a vertical red line: results found in this region are numerically indistinguishable to the zero mode. Finally we note that the fraction of observations corresponding to the zero mode is progressively reduced as we increase the size of the system: this a consequence of the fact that the number of unstable modes with finite frequencies grows as a function of system size while the zero mode remains one (or, within numerical precision, a single pair of complex conjugate eigenvalues). We thus expect that in the limit of large systems, the overwhelming majority of unstable modes have finite time-scales.

\bibliography{NeutModes}
\end{document}